\documentclass[aip,apl,reprint,twocolumn]{revtex4-1}

\usepackage{graphicx}
\usepackage{dcolumn}
\usepackage{bm}
\usepackage{xfrac}
\usepackage{SIunits}
\usepackage{physics}
\usepackage{color}
\usepackage[usenames,dvipsnames]{xcolor}
\usepackage{hyperref}
\usepackage{footmisc}
\usepackage{amsmath}
\usepackage{ulem}
\usepackage{natbib}
\usepackage[utf8]{inputenc}
\usepackage[T1]{fontenc}
\usepackage{mathptmx}

\newcommand{\wc}{\omega_{\mathrm{res}}}

\newcommand{\wL}{w_{\mathrm{L}}}
\newcommand{\QL}{Q_{\mathrm{L}}}

\newcommand{\geff}{g_{\mathrm{eff}}}
\newcommand{\keff}{\kappa_{\mathrm{eff}}}

\begin{document}

\title{Maser Threshold Characterization by Resonator Q-Factor Tuning}

\author{Christoph W. Zollitsch}
\email{c.zollitsch@ucl.ac.uk}
\affiliation{Department of Chemistry, Saarland University, Saarbr\"ucken, 66123, Germany}
\affiliation{Department of Physics \& Astronomy, University College London, Gower Street, WC1E 6BT, UK}
\affiliation{London Centre for Nanotechnology, University College London, 17-19 Gordon Street, WC1H 0AH, UK}
\author{Stefan Ruloff}
\affiliation{Department of Chemistry, Saarland University, Saarbr\"ucken, 66123, Germany}
\author{Yan Fett}
\affiliation{Department of Chemistry, Saarland University, Saarbr\"ucken, 66123, Germany}
\author{Haakon T. A. Wiedemann}
\affiliation{Department of Chemistry, Saarland University, Saarbr\"ucken, 66123, Germany}
\author{Rudolf Richter}
\affiliation{Department of Chemistry, Saarland University, Saarbr\"ucken, 66123, Germany}
\author{Jonathan D. Breeze}
\affiliation{Department of Physics \& Astronomy, University College London, Gower Street, WC1E 6BT, UK}
\author{Christopher W. M. Kay}
\email{christopher.kay@uni-saarland.de}
\affiliation{Department of Chemistry, Saarland University, Saarbr\"ucken, 66123, Germany}
\affiliation{London Centre for Nanotechnology, University College London, 17-19 Gordon Street, WC1H 0AH, UK}

\date{\today}

\maketitle

\section*{Abstract}
\textbf{Whereas the laser is nowadays an ubiquitous technology, applications for its microwave analogue, the maser, remain highly specialized, despite the excellent low-noise microwave amplification properties. The widespread application of masers is typically limited by the need of cryogenic temperatures. The recent realization of a continuous-wave room-temperature maser, using NV$^-$ centers in diamond, is a first step towards establishing the maser as a potential platform for microwave research and development, yet its design is far from optimal. Here, we design and construct an optimized setup able to characterize the operating space of a maser using NV$^-$ centers. We focus on the interplay of two key parameters for emission of microwave photons: the quality factor of the microwave resonator and the degree of spin level-inversion. We characterize the performance of the maser as a function of these two parameters, identifying the parameter space of operation and highlighting the requirements for maximal continuous microwave emission.}
\begin{figure}[b!]
 \includegraphics[]{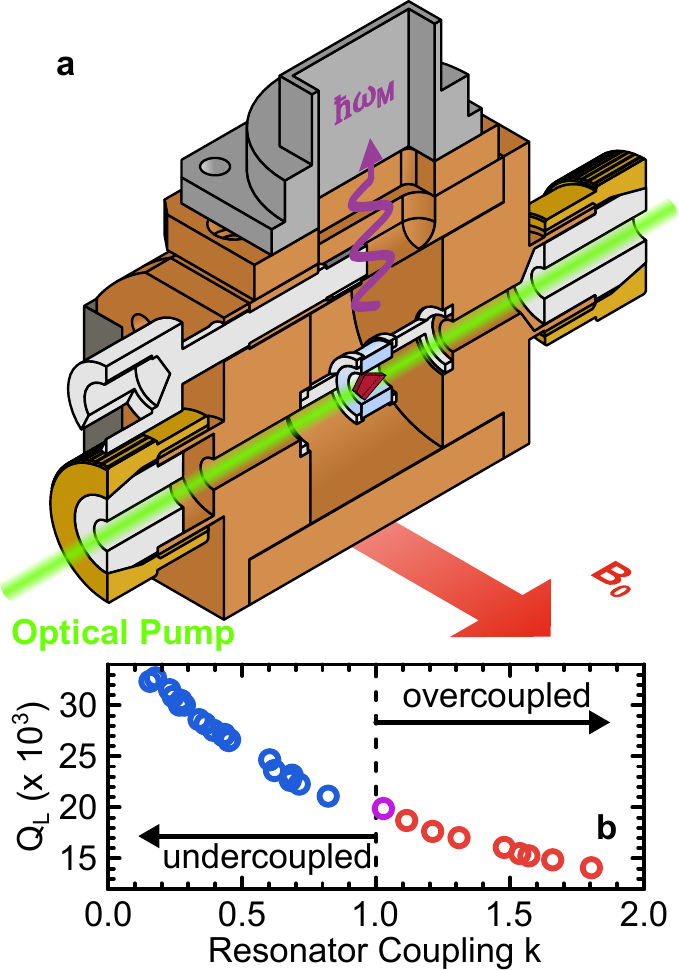}
 \caption{\textbf{Schematic cross-section of the microwave cavity-resonator setup.} \textbf{a} Microwave cavity, holding the sapphire ring resonator. The schematic shows half of the cavity to reveal the interior. The light blue sapphire ring is held by two Teflon holders. On the top is the iris coupled waveguide port. The coupling to the resonator is adjusted by a Teflon screw with a metal ring at its tip. The $532\,\nano\meter$ optical pump laser is aligned along the symmetry axis of the cylindrical resonator, while the static magnetic field $B_0$ is perpendicular to it. Finally, the diamond sample is placed inside the resonator. \textbf{b} Loaded $Q$-factor as a function of the resonator coupling, showing the full range of coupling, with the diamond sample in the resonator.
 \label{fig01}}
\end{figure}

\section*{Introduction}
The first maser system was realized using ammonia molecules in the gas phase \cite{Gordon1955} and applications for signal amplifiers, frequency standards or as spectrometers were subsequently proposed. However, the need for cryogenic and/or high vacuum environments restricted miniaturization and integration towards more general applications. Maser research and development was mainly focused on low-noise microwave receiving systems for deep-space antenna networks \cite{Reid1973}, and several other maser systems based on ruby \cite{Makhov1958}, atomic hydrogen \cite{Goldenberg1960} or Rydberg atoms \cite{Moi1983} were realized. These systems were all still subject to the same restrictions and the focus on fundamental research into masers declined. The field was reinvigorated upon the realization of a room-temperature pulsed maser in an optically pumped crystal of pentacene-doped \textit{p}-terphenyl, placed inside a high quality factor microwave resonator \cite{Oxborrow_Nature2012, Breeze_NComms2015, Salvadori2017}. Here, masing was achieved not only without the need of cryogenics or a high-vacuum environment, but also with easily accessible optical pump rates. Shortly afterwards, a proposal for a continuous-wave room-temperature maser in optically spin polarized, negatively charged nitrogen vacancy (NV$^-$) centers in diamond \cite{Jin_NComms2015} and its consecutive experimental realization \cite{Breeze_Nature2018} followed. The excellent low-noise amplification properties of the maser have been demonstrated in recent work \cite{Sherman2022, Koppenhofer2022, Jiang2022_amp} and has found application in enhanced quantum sensing of molecular spin ensembles \cite{Wu2022}. The search for other solid-state maser materials, such as SiC \cite{Gottscholl2022} has continued, and an application in quantum technology by creating a maser system based on Floquet states in Xe atoms \cite{Jiang2022} has also been reported. 

With the increased focus on room-temperature solid-state masers a quantitative experimental characterization of the parameter space of operation supplements the current endeavours to optimize the performance of such systems. Understanding the behaviour of the maser performance, defined by the level-inversion of the spin ensemble and the loaded $Q$-factor $\QL$ of the resonator, can lead to higher output power maser systems.
To guarantee reproducibility and maximise the power output of maser-based technologies, a complete understanding of the system parameter space, including the minimal requirements to surpass the masing threshold, is required.

Here, we present an experimental setup that we exploited to investigate the maser performance of a NV$^-$ spin ensemble hosted in diamond as a function of the resonator quality factor and the degree of spin level-inversion. In the resulting maser threshold diagram, we can clearly identify the threshold for maser action, thereby obtaining a set of experimental boundary conditions for the optimal operation of a maser system. Additionally, our optimized setup yields the highest continuous-wave maser output power reported to date.

\section*{Results and Discussion}
\subparagraph{Experimental setup:} We use a cylindrical dielectric ring resonator made of sapphire to deliver/detect resonant microwaves to/from the NV$^-$ centers contained in the diamond host. A key parameter for the continuous maser emission is the loaded $Q$-factor, $\QL$, of the resonator, which is defined by $\sfrac{1}{\QL} = \sfrac{1}{Q_{\mathrm{int}}} + \sfrac{1}{Q_{\mathrm{ext}}}$ with the internal $Q$-factor, $Q_{\mathrm{int}}$, and the external $Q$-factor, $Q_{\mathrm{ext}}$. Although sapphire dielectric resonators exhibit low dielectric losses \cite{Shtin2009}, radiative losses typically dominate and prevent high internal quality factors. Hence, to suppress radiative losses the resonator is placed inside a metal cavity. The cavity design is further constrained by two conditions: (i) it has to fit between the poles of our electromagnet system, which gives the static magnetic field $B_0$ used to tune the energy levels of the NV$^-$ spins via the Zeeman interaction, and (ii) the resonance frequency of the cavity containing the sapphire ring is required to be within the $9 - 10\ \giga\hertz$ (X-band) frequency range, to allow a fast pre-characterization of the NV$^-$ spin ensemble by conventional electron spin resonance (ESR). To this end, we designed a cylindrical cavity made of oxygen-free high-thermal conductivity (OFHC) copper, plated with thin layers of (first) silver and (second) gold to prevent oxidation of the metal surfaces, thereby minimising additional resistive losses.  

\begin{figure}[t]
 \includegraphics[]{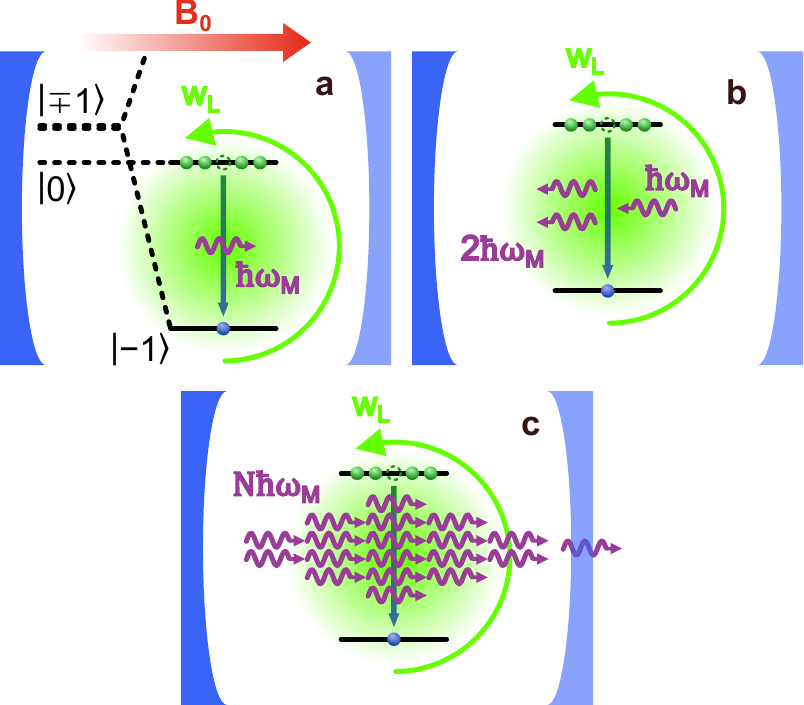}
 \caption{\textbf{Schematic of the maser emission process.} \textbf{a - c} The blue mirrors are the optical analogue of a high-$Q$ microwave resonator, with one mirror being fully reflective (dark blue) and one weakly transmissive (light blue); \textbf{a} Energy states of the NV$^-$ spin ensemble aligned along $B_0$. The $\ket{-1}$ state is tuned below the $\ket{0}$ state by the Zeeman interaction, while the optical pump generates a population inversion and spins spontaneously emit microwaves. \textbf{b} Spontaneously emitted or thermally stimulated microwaves can stimulate the emission of additional microwaves. \textbf{c} For sufficiently large optical pump rates and a high $\QL$ coherent microwave emission is established.
 \label{fig02}}
\end{figure}
Figure \ref{fig01}\,(a) shows a 3D schematic of the fully assembled cavity cut at a symmetry plane to reveal the interior. The cavity has sample entries at opposite ends of the cylinder symmetry axis for sample and optical access. In the center of the cavity the resonator is held in place by two wire-frame Teflon holders. The holders are machined to have minimal volume in order to minimize additional dielectric losses. Microwave power is coupled in and out via a single waveguide iris port on the top. This port represents the external quality factor $Q_{\mathrm{ext}}$ and together with $Q_{\mathrm{int}}$ the resonator coupling $k = \sfrac{Q_{\mathrm{int}}}{Q_{\mathrm{ext}}}$ is defined. $k$ is controlled via a Teflon screw with a metal ring at its tip. Changing the coverage of the iris by the metal ring allows a continuous change \cite{Accatino1994} from over-coupled to under-coupled: the regimes where $\QL$ is dominated by external losses or intrinsic losses, respectively. Figure \ref{fig01}\,(b) shows the achievable $\QL$ as a function of $k$ for our resonator-cavity system, loaded with the diamond sample. 
The resonator parameters are extracted from microwave reflection measurements, using a vector network analyzer. By fitting the microwave reflection as a function of frequency with a Lorentzian model function the parameters for resonance frequency, resonator coupling and internal quality factor are determined. A description of the model function can be found in the methods section.
We use the TE$_{01\delta}$ mode for our experiments, where the electric field is mostly contained in the sapphire ring and the magnetic field is mostly focused in the bore. Without a sample, the sapphire resonator has a resonance frequency $\wc/2\pi = 9.25\,\giga\hertz$ and the $\QL = 42,500$ when fully under-coupled. We define the resonator as fully under-coupled when $k = 0.0027$, where the Iris is no longer covered by the metal tip of the Teflon screw. Here, $\QL$ remains unchanged upon further extraction of the screw. For detailed cavity and resonator dimensions see Supplementary Note 4 \cite{supplement}.

\subparagraph{Maser working principles:} The process of continuous emission of microwave photons from the NV$^-$ centers is schematically shown in Fig.\,\ref{fig02}\,(a) to (c). The diamond hosting the NV$^-$ centers is placed inside a high-$Q$ resonator, which is highly under-coupled ($k \ll 1$). The resonator may be pictured in analogue to its optical counterpart the laser, with one perfectly reflective mirror and one weakly transmitting mirror. The latter represents the iris coupled single microwave port on the microwave cavity. 

The applied static magnetic field $B_0$ lifts the degeneracy of the $\ket{\pm\,1}$ states and, tunes the $\ket{-1}$ state energetically below the $\ket{0}$ state such that the splitting $\hbar\omega_\mathrm{M}$ is resonant with the microwave resonator frequency. The experiment is performed only on one sub-set of NV$^-$ centers which are aligned with the external magnetic field $B_0$. This provides the shown level structure, having the largest Zeeman splitting of the energy levels and consequently the largest initial population difference at Boltzmann equilibrium. By illuminating the NV$^-$ centers continuously with a $532\,\nano\meter$ laser, the spin populations which are initially at Boltzmann equilibrium are predominately pumped into the $\ket{0}$ state \cite{Robledo_NJP2011}, resulting in a level-inversion (see Fig.\,\ref{fig02}\,(a)). A description of the optical spin polarization process is found in Supplementary Note 2 \cite{supplement}. Finally, the laser polarization is required to be aligned along the NV$^-$ defect axis to achieve most efficient pump rates \cite{Kai-Mei2009, Doherty2013}.

To trigger a collective stimulated emission, an initial photon with $\hbar\omega_\mathrm{M}$ is required. This is provided either by an externally applied seeding photon, due to spontaneous emission or thermal photons (see Fig.\,\ref{fig02}\,(b)). From this point an avalanche of stimulated photons is created, forming a coherent microwave field inside the resonator (see Fig.\,\ref{fig02}\,(c)). If the laser pump rate is sufficient to maintain the level inversion and the resonator loaded quality factor is high enough to support a large enough coherent microwave field, continuous microwave emission is achieved.
\begin{figure}[b]
 \includegraphics[]{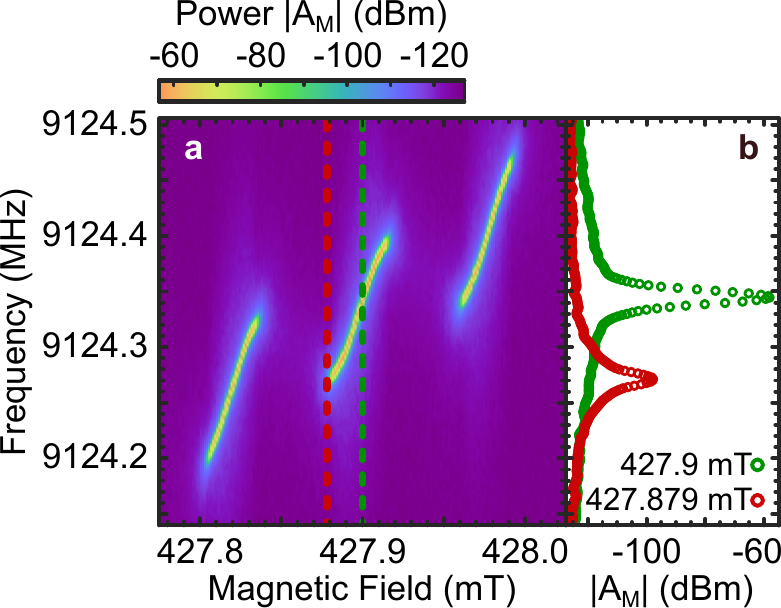}
 \caption{\textbf{Maser emission spectrum.} \textbf{a} Maser emission power $\left| A_\mathrm{M} \right|$, in logarithmic units, as a function of frequency and static magnetic field at $\wL = 430\,\second^{-1}$ and $\QL$ of about $33,500$. \textbf{b} $\left| A_\mathrm{M} \right|$ as a function of frequency for two different magnetic field values, indicated by the red and green dashed lines in \textbf{a}.
 \label{fig03}}
\end{figure}

\subparagraph{NV defect axis alignment:} In our experimental setup, the diamond sample is held inside a quartz ESR tube, supported between two additional quartz tubes fitted inside the first one. The tube is inserted into the cavity and positioned such that the diamond is located at the center of the sapphire ring where the magnetic component of the microwave field is largest. The cavity with sample is mounted between the poles of a electromagnet system. We connect either a conventional ESR spectrometer, a vector network analyzer or a spectrum analyzer to the microwave port of the cavity, to perform low-power microwave spectroscopy of the NV$^-$ spin transitions or to study the maser emission. A goniometer is attached to the sample tube, allowing a precise rotation of the quartz tube containing the diamond with respect to the static magnetic field. 
Conventional ESR as a function of $B_0$ and rotation angle is performed to find an orientation of the diamond where the defect axis of a sub-set of NV$^-$ centers is mostly parallel to the applied magnetic field. For such an orientation the NV$^-$ spins feature an energy level scheme of the electronic ground state as schematically shown in Fig.\,\ref{fig02}\,(a) and the states $\ket{0}$ and $\ket{\pm1}$ can be considered pure. This can be characterized by the frequency/magnetic field splitting between the low-field $\left( \ket{0}\,\mapsto\,\ket{+1} \right)$ and high-field $\left( \ket{0}\,\mapsto\,\ket{-1} \right)$ transitions corresponding to twice the zero-field splitting $D$. Away from this alignment the states become mixed, resulting in a smaller splitting than $2D$ between the two allowed transitions \cite{Mrozek2015, Jeong2017}  and a smaller maximal achievable spin polarization. Without laser illumination, we find a maximal splitting of about $205\,\milli\tesla$ or $5.762\,\giga\hertz$, agreeing well with twice the zero-field splitting, $D$, of NV$^-$ centers in diamond.

\subparagraph{Maser emission spectrum characterization:} Having optimized the orientation of the NV$^-$ centers, we characterize the performance of our maser setup by analyzing the microwave emission spectrum. Figure\,\ref{fig03}\,(a) shows the color encoded maser emission power $\left| A_\mathrm{M} \right|$ as a function of frequency and static magnetic field for a laser pump rate $\wL = 430\,\second^{-1}$ and with the resonator fully under-coupled, $\QL \approx Q_\mathrm{int}$, which gives a loaded quality factor $\QL$ of $33,500$.
The three bright lines represent the maser emission of the three $^{14}$N hyperfine transitions of the NV$^-$ centers oriented along $B_0$. In this configuration, we achieve a maximum maser emission power of $-56.5\,\deci\bel\milli$.
The maser emission shows a finite frequency-magnetic field dispersion, where the maser power is maximal in the center of the line. This is illustrated in Fig.\,\ref{fig03}\,(b) which depicts the maser emission power as a function of frequency for two fixed magnetic fields. The dispersion results from the hybridization of the microwave resonator mode and the resonant NV$^-$ transition. The resonator resonance frequency $\wc/2\pi = 9.12\,\giga\hertz$ lies at the center of the middle maser emission line. The inverted spin population causes a dispersive shift of the resonator frequency to lower frequencies for the lower magnetic field emission line and respectively to a shift to higher frequencies for the higher magnetic field emission line. The maser dispersion can be described via a Tavis-Cummings model with an inverted spin polarisation, where the frequency range covered by the maser emission lines increases with increasing inversion.

\begin{figure}[b]
 \includegraphics[]{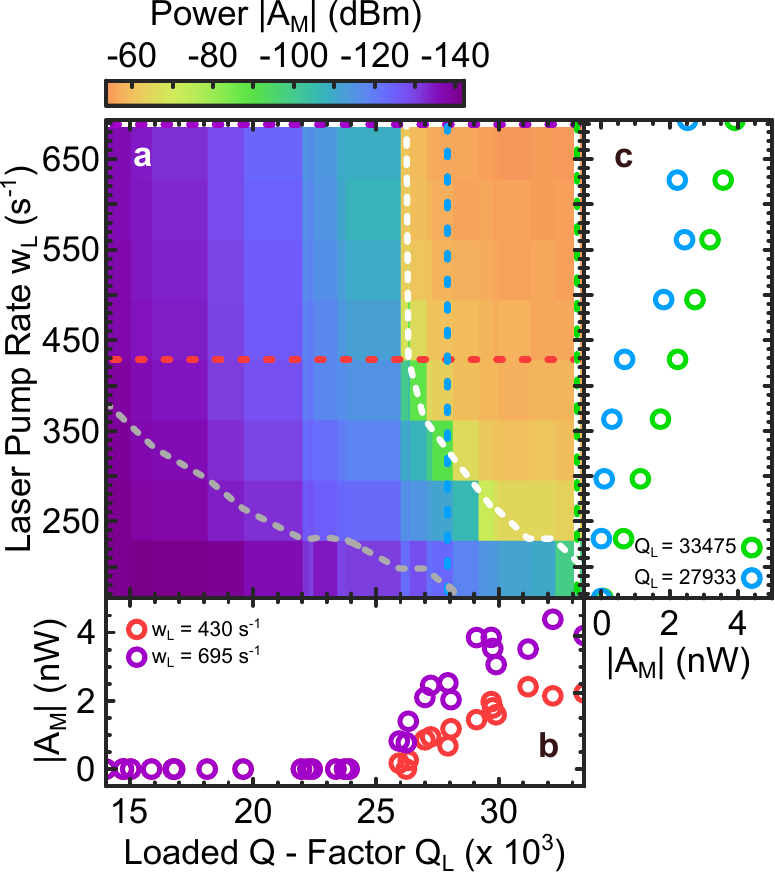}
 \caption{\textbf{Maser threshold diagram.} \textbf{a} Maximal maser emission power $\left| A_\mathrm{M} \right|$, in logarithmic units, of the central hyperfine maser transition line as a function of laser pump rate $\wL$ and the loaded quality factor $\QL$ of the resonator. The white and grey dashed lines represent the theoretical threshold for masing, with and without temperature effects, respectively. \textbf{b} $\left| A_\mathrm{M} \right|$ as a function of $\QL$ for two $\wL$, indicated by the purple and red dashed lines in \textbf{a}. \textbf{c} $\left| A_\mathrm{M} \right|$ as a function of $\wL$ at two $\QL$, indicated by the blue and green dashed lines in \textbf{a}.
 \label{fig04}}
\end{figure}

In order to determine the dependency of the maser output on the laser pump rate $\wL$ and the loaded quality factor $\QL$, both were varied systematically. The latter is dependent on the resonator coupling $k$ which is controlled by the iris screw. Figure \ref{fig04}\,(a) shows the peak maser power $\left| A_\mathrm{M} \right|$ of the central maser line as a function of $\wL$ (from $165\,\second^{-1}$ to $695\,\second^{-1}$) and $\QL$ (from $14,000$ to $33,500$). Note, that after each change of $\wL$ we wait 45\,min before starting measurements to allow the diamond to reach a thermal equilibrium (see Supplementary Note 1 for details \cite{supplement}). For low loaded quality factors, i.e. an over-coupled resonator, there is no microwave emission for all laser pump rates studied and the signal amplitude is represented by the noise floor of the spectrum analyzer. Note that critical resonator coupling ($k = 1$) is achieved at $\QL \approx 20,000$, marking the transition between over- and under-coupling \cite{Goeppl2008}. With increasing $\QL$ a weak microwave emission is observed for moderate to high laser pump rates. Here, the emission spectrum is broad and governed by the amplification of thermal photons residing in the resonator \cite{Jin_NComms2015}. In this region, the resonator losses are still too high to allow the build-up of sufficient stimulated photons for continuous masing. For $\QL > 26,000$ and $\wL > 400\,\second^{-1}$ the rate of stimulated emission exceeds the losses in the resonator and the spin system, and continuous masing is established. Figure\,\ref{fig04}\,(b) shows the evolution of the peak maser power as a function of the loaded quality factor for two different laser pump rates. Upon passing the threshold $\QL$ of $26,000$, $\left| A_\mathrm{M} \right|$ increases until reaching an onset of saturation for the highest $\QL$, where the final level of saturation depends on the laser pump rate. A $\QL$ larger than $26,000$ further promotes the stimulated microwave emission, thus reducing the threshold laser pump rate required. For a fully under-coupled resonator with $\QL = 33,500$, continuous microwave emission is already achieved for $\wL > 200\,\second^{-1}$. Figure \ref{fig04}\,(c) shows the peak maser power $\left| A_\mathrm{M} \right|$ as a function of the laser pump rate $\wL$ at two different loaded quality factors of the resonator. The peak maser power increases with increasing laser pumping, reflecting the achieved level-inversion. Consequently, for the largest $\QL$ and highest $\wL$, which guarantees a stable microwave field inside the resonator and the largest level inversion, respectively, the highest maser power of $-54.1\,\deci\bel\milli$ is achieved. This is an improvement of more than three orders of magnitude in maser power over the initial report on a NV$^-$-based maser in Ref.\,\onlinecite{Breeze_Nature2018} with reported $-90.3\,\deci\bel\milli$.
The threshold laser pump rate $w_{\mathrm{th}}$ for masing can be derived from the steady-state Heisberg equations of motion based on the coupled resonator-spin ensemble Tavis-Cummings model, including dissipative channels for the resonator and spins, and is given by \cite{Breeze_Nature2018, Jin_NComms2015, Kolobov1993}

\begin{equation}
    w_{\mathrm{th}} = \frac{\eta}{T_1 \left( \frac{g_0^2 N}{\kappa_0 \gamma} - 1 \right)}.
    \label{eq:threshold}
\end{equation}
Here, $T_1$ is the effective longitudinal relaxation time of the NV$^-$ spins, $g_0$ gives the strength of the magnetic dipole coupling between a single spin and a single microwave photon \cite{Haroche2013}, $N$ is the absolute number of spins per hyperfine transition and per NV$^-$ defect axis and $\kappa_0$ and $\gamma$ are the resonator and spin loss rates (HWHM), respectively. We estimate $g_0/2\pi$ by finite element simulations of the magnetic field profile inside the sapphire ring resonator\cite{Oskooi2010} to an average value of $0.244\,\hertz$ and determine $\gamma/2\pi$ and $N$ via low-power microwave spectroscopy of the low-field transition $\left( \ket{0}\,\mapsto\,\ket{+1} \right)$ to $530.8\,\kilo\hertz \pm 7.7\,\kilo\hertz$ and $2.32\,\times\,10^{13}$, respectively. A detailed derivation of these system parameters is found in Supplementary Note 3 \cite{supplement}. The resonator loss rate is defined through $\QL$ as $\kappa_0 = \sfrac{\wc}{2\QL}$, which we control by changing the resonator coupling. The scaling factor $\eta$ modifies the optical pump rate defined for a two-level system within the framework of the Tavis-Cummings model to take all seven energy levels involved in the pumping of a single NV$^-$ spin into account. We extract $\eta = 14.05$ from the calculated spin-level inversion as a function of $\wL$ \cite{Jin_NComms2015, Breeze_Nature2018}, by solving the set of optical pump rate equations in the steady-state \cite{Sherman2021}. For $T_1$ we explicitly take the influence of an optical pump into account. In addition to polarizing spins, the pump leads to excessive heating of the diamond, which decreases the $T_1$ time \cite{Jarmola2012}. We determine $T_1$ for each $\wL$, finding $5.2\,\milli\second$ for low $\wL$ and a minimum of $1.5\,\milli\second$ for the highest $\wL$, where a detailed description of the $T_1$ dependence on $\wL$ is given in Supplementary Note 1 \cite{supplement}. The white dashed line in Fig.\,\ref{fig04}\,(a) gives the masing threshold $w_\mathrm{th}$ as a function of the loaded quality factor and is in excellent agreement with the experimentally found threshold of our maser, when acknowledging the reduction of the relaxation time due to high pump rates. For comparison, we include the expected threshold for a fixed $T_1 = 5.2\,\milli\second$ (grey dashed line), demonstrating the significant influence of laser heating.

\section*{Conclusion}
To summarize, our characterization setup allows the resonator coupling to be continuously and precisely adjusted between over- and under-coupled in conjunction with the optical pump rate. This permits the detailed study of the performance of the maser as a function of the rate of stimulated emission and the degree of level-inversion. Control of these parameters enabled the first experimental verification of the maser threshold equation over a wide parameter space. Thus, the regions of microwave emission below the detection limit, thermal photon amplification and continuous masing could be identified in a NV$^-$ diamond maser.

Our results highlight an efficient operation of the maser is in the highly under-coupled regime. Clearly, this limits the maser output power as only a small fraction of the microwaves in the weakly coupled resonator can exit. For applications, a higher output is essential and, hence, a larger coupling to the resonator is required. The resulting threshold diagram suggests that either higher pump rates or a resonator with a higher $Q_{\mathrm{int}}$ will enable a larger coupling of the resonator. However, higher pump rates not only require bulky laser systems but cause the sample to heat up. We demonstrate that the $T_1$ relaxation time is shortened, thus reducing the spin inversion. Indeed, it is clear for our results that pump rates can be reduced to a level where small (e.g. $< 1\,\watt$) laser diodes can be employed instead. Therefore, increasing the $Q$ factor of the resonator is a more viable approach, although it is necessary to point out that this will decrease the bandwidth\cite{Sherman2022}.
Finally, the optimized design yielded an increase of the maximal maser output of more than three orders of magnitude compared to the initial report on a NV$^-$-based maser from $-90.3\,\deci\bel\milli$ to $-54.1\,\deci\bel\milli$ \cite{Breeze_Nature2018}. We attribute this mainly to an improved heat management of the diamond sample to limit a reduction in $T_1$ time and consequently limit the increase of the masing threshold. Thus it provides not only a blueprint for solid-state based maser systems and but also sets a benchmark for future characterization and optimization studies.

\section*{Methods}
\textbf{Diamond Sample:} Our diamond sample is of rhombic shape with its long axis having about $5\,\milli\meter$, its short axis having about $4\,\milli\meter$ and a thickness of $1\,\milli\meter$. The diamond consists of natural abundance carbon and we estimate the total number of NV$^-$ to $2.78\,\times\,10^{14}$ or $0.16\,$ppm (see Supplementary Note 3 for details on the number of spins estimate \cite{supplement}). At room temperature and no optical pump the NV$^-$ feature a $T_2 = 25\,\micro\second$, determined by pulsed ESR.

\textbf{Laser Pumping:} We use a $532\,\nano\meter$ \textit{Coherent Verdi V-5} laser to optically pump the spin population of the NV$^-$ centers. The laser features a spot size of about $4\,\milli\meter$ and hits the diamond sample on its flat edge, an area of about $4\,\milli\meter \times 1\,\milli\meter$. We determine the laser pump rate as $\wL = \sfrac{\sigma P_{\mathrm{pump}}}{A_{\mathrm{pump}}\hbar\omega_{\mathrm{pump}}}$ (Ref.\,\onlinecite{Breeze_Nature2018}), with the one-photon absorption cross-section for NV$^-$ centers $\sigma = 3.1 \times 10^-{21}\,\meter^2$, the pump laser power $P_{\mathrm{pump}}$, the laser spot area $A_{\mathrm{pump}}$ and the laser frequency $\omega_{\mathrm{pump}}$.

\textbf{Microwave Spectroscopy:} We study the microwave emission from the maser with a Keysight N9020B MXA $10\,\hertz$ to $44\,\giga\hertz$ spectrum analyzer. To improve the SNR we pre-amplify the maser signal, using a Mini Circuits Low Noise Amplifier ZX60-06183LN+ prior detection by the spectrum analyzer. The low power microwave spectrocopy was carried out using an Anritsu vector network analyzer MS46122B $1\,\mega\hertz$ to $43.5\,\giga\hertz$. For conventional ESR measurements a Bruker EMXplus spectrometer is used.

\textbf{Resonator Coupling:} We determine the microwave resonator coupling $k$ by measuring the microwave reflection as a function of frequency with a VNA, far detuned from the NV$^-$ centers. We fit the magnitude of the reflection scattering parameter $\left| S_{11} \right|$ of the resonator for different positions of the Iris screw, using a Lorentzian model function derived from an equivalent circuit model, which depends on $k$ \cite{Wang2018}.
\begin{equation}
    \left| S_{11} \right| = A \left| \frac{k - 1 - iQ_\mathrm{int}\left( \frac{\omega}{\omega_\mathrm{res}} - \frac{\omega_\mathrm{res}}{\omega} \right)}{k + 1 + iQ_\mathrm{int}\left( \frac{\omega}{\omega_\mathrm{res}} - \frac{\omega_\mathrm{res}}{\omega} \right)} \right|.
    \label{eq:lorentz}
\end{equation}
In addition, we can extract the internal quality factor $Q_{\mathrm{int}}$ and via the relation for the resonator coupling $k = Q_\mathrm{int} / Q_\mathrm{ext}$ the external Q-factor $Q_\mathrm{ext}$ and hence the loaded quality factor $\QL$, as well as the resonator frequency $\wc$ and its loss rate $\kappa_0$.

\section*{Data Availability}

The data that supports the findings of this study is available upon reasonable request. 

\section*{Author Contribution}
S.R., R.R. and C.W.M.K. designed and optimized the resonator. The experimental setup and its automation was realized by S.R., Y.F., H.T.A.W., C.W.Z. and C.W.M.K.. C.W.Z. performed the experiments and the data analysis with input from J.D.B. and C.W.M.K.. C.W.Z. supported by C.W.M.K. wrote the manuscript with input from all authors. C.W.Z, J.D.B. and C.W.M.K. conceived the study.

\section*{Acknowledgment}
This study is supported by EPSRC through EP/S000690/1 and EP/S000798/2 and the Royal Society through URF$\backslash$R1$\backslash$191297.
We greatly acknowledge the work and expertise of the workshop personnel Stefanie Porger and Jens Wiegert at the University of Saarland for building the cavity and components for it. 

\section*{Competing interests}

The Authors declare no conflict of interests.

\section*{References}

\setcounter{equation}{0}
\setcounter{figure}{0}
\setcounter{table}{0}
\setcounter{page}{1}
\makeatletter
\pagebreak
\widetext
\begin{center}
\textbf{\large Supplementary Material: Maser Threshold Characterization by Resonator Q-Factor Tuning}
\end{center}

\setcounter{equation}{0}
\setcounter{figure}{0}
\setcounter{table}{0}
\setcounter{page}{1}
\makeatletter

\renewcommand{\figurename}{Supplementary Figure }
\renewcommand{\theequation}{S\arabic{equation}}
\renewcommand{\bibnumfmt}[1]{[S#1]}
\renewcommand{\citenumfont}[1]{S#1}

\section*{Supplementary Note 1: Temperature Increase of the Diamond due to Optical Pumping}

The study of the maser output power as a function of optical pump rate requires considerable laser powers. The highest pump rate $\wL = 695\,\second^{-1}$ is achieved by a laser power of $1.05\,\watt$. Such high powers lead to excess heating of the diamond, which cannot be dissipated fully in our measurement setup. As described in the main text, the diamond sample is held inside a quartz ESR tube ($5\,\milli\meter$ outer diameter (OD)), supported between two additional quartz tubes ($4\,\milli\meter$ OD) fitted tightly inside the first one. Through the ESR tubes the diamond is indirectly thermally connected to the sapphire ring as well as the tube holders on the cavity body, which allows a finite thermal transport of the excess heating from the laser. In addition, we introduce a weak nitrogen gas flow into the cavity body to achieve an additional cooling effect. The nitrogen gas is connected to an inlet on the waveguide and flows via the Iris into the cavity and out via gaps in the ESR tube holders. Together, a finite thermal equilibrium in the diamond will establish after some time. For the experiments presented in the main text, we wait 45 minutes before a start of a measurement, after changing the laser power.
The longitudinal relaxation time $T_1$ exhibits a strong temperature dependence \cite{Jarmola2012}. For temperatures above room-temperature the relaxation rate $\sfrac{1}{T_1}$ predominately increases with $T^5$. As the maser threshold $w_{\mathrm{th}}$ is indirect proportional to $T_1$ (see main text), the threshold will increase due to the effects of laser heating.

\begin{figure}[b]
 \includegraphics[]{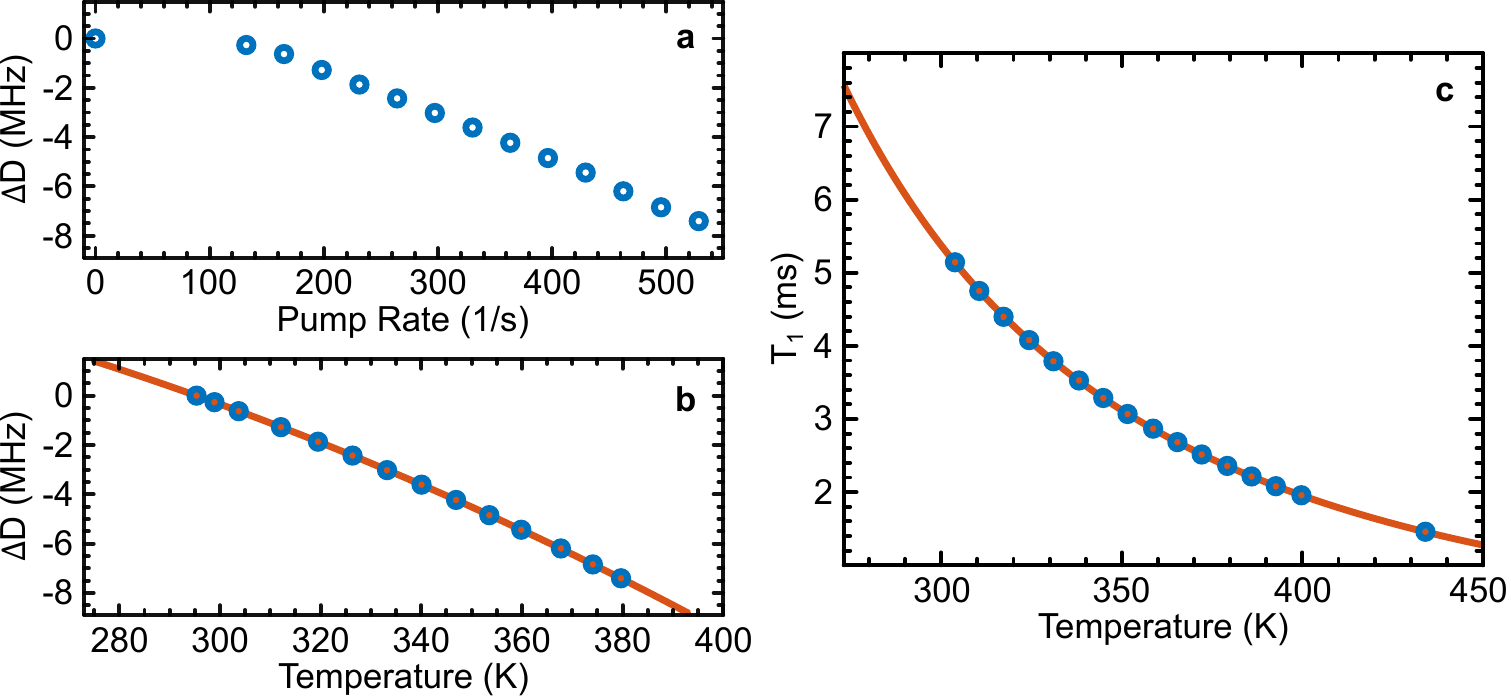}
 \caption{\textbf{Temperature dependence of zero-field splitting and relaxation time.} \textbf{a} Color encoded microwave reflection $\left| S_{11} \right|$ as a function of VNA frequency and static magnetic field, at a pump rate of $695\,\second^{-1}$. \textbf{b} Effective loss rate $\kappa_{\mathrm{eff}}$ of the resonator - spin ensemble hybrid system as a function of the static magnetic field. The orange solid line is a fit of Eq.\,(\ref{eq:keff}) to the data.
 \label{S01}}
\end{figure}
To determine the temperature increase in our maser characterization setup, we first determine the change of the zero-field splitting $\Delta D$ as a function of optical pump rate, via low power microwave spectroscopy of the low- and high-field NV$^-$ spin transitions, and is shown in Supplementary Figure\,\ref{S01}\,(a). With increasing pump rate the zero-field splitting is decreasing. We define the room-temperature value of $D = 2.88\,\giga\hertz$ as the reference for the relative shift $\Delta D$. Due to this temperature dependence, NV centers are highly sensitive thermometers. The temperature dependence can be described by a second order polynomial \cite{Ouyang2023},
\begin{equation}
    \Delta D = \Delta D_\mathrm{cal} + a_1 T + a_2 T^2,
\end{equation}
with coefficients $a_1 = 3.279\,\times\,10^{-2}\,\mega\hertz\kelvin^{-1}$ and $a_2 = -1.787\,\times\,10^{-4}\,\mega\hertz\kelvin^{-2}$. With the constant offset $\Delta D_\mathrm{cal} = 5.9\,\mega\hertz$ we calibrate the model to produce a zero shift at room-temperature. With the model function we can relate the measured $\Delta D$ to a temperature, as shown in Supplementary Figure\,\ref{S01}\,(b).

Finally, we can find the corresponding $T_{1,\mathrm{eff}}$ times for each optical pump rate used in the maser characterization. Supplementary Figure\,\ref{S01}\,(c) shows the calculated temperature dependence of $T_1$ (solid red line) together with the $T_{1,\mathrm{eff}}$ for each pump rate used in the experiment (symbols), ranging from $165\,\second^{-1}$ to $695\,\second^{-1}$. We calculate the relaxation time following the formalism presented in \onlinecite{Jarmola2012}
\begin{equation}
    T_1 = \left( A_1 + \frac{A_2}{e^{\sfrac{\Delta}{k_\mathrm{b}T}} - 1} + A_3 T^5 \right)^{-1}.
\end{equation}
The coefficients are $A_1 = 0.06\,\second^{-1}$, $A_2 = 2.1\,\times\,10^3\,\second^{-1}$ and $A_3 = 2.2\,\times\,10^{-11}\,\second^{-1}\kelvin^{-5}$, $\Delta = 1.17\,\times\,10^{-20}\,\joule$ is the dominant local vibrational energy and $k_\mathrm{b}$ is the Boltzman constant.

\section*{Supplementary Note 2: Optical Spin Polarization of the NV$^-$ Spin Ensemble}

\begin{figure}[t]
 \includegraphics[]{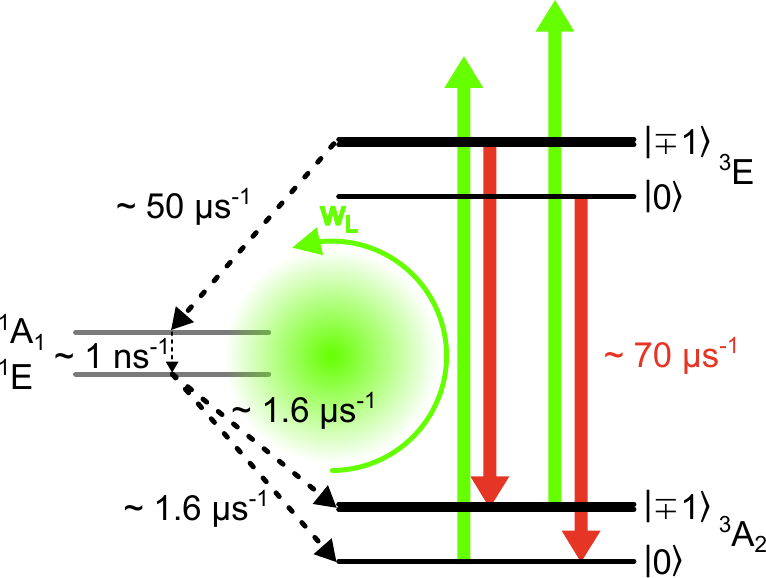}
 \caption{\textbf{Optical spin polarization scheme.} Schematic of the electronic energy levels of a NV$^-$ spin ensemble. The triplet states $^3$A$_2$ and $^3$E are considered to be quantized along a principal axis of the zero-field splitting tensor. A continuous optical pump via a $532\,\nano\meter$ laser can achieve a majority population of the $^3$A$_2$ $\ket{0}$ state, due to the spin selective intersystem crossing in the excited $^3$E triplet.
 \label{fig_S_opticla_pump}}
\end{figure}

Supplementary Figure \ref{fig_S_opticla_pump} shows the schematic electronic energy level scheme for NV$^-$ spins at zero applied magnetic field \cite{Robledo_NJP2011, Breeze_Nature2018}. We use a green $532\,\nano\meter$ laser to excite the population from the electronic ground state $^3$A$_2$ into the excited state $^3$E. Hereby, all of the spin triplet states $\ket{0}$ and $\ket{\pm1}$ being pumped equally. The excited triplet state decays via a red fluorescence back in the $^3$A$_2$. Additionally, the electrons from the excited $\ket{\pm1}$ states undergo a spin selective intersystem crossing via the singlet states $^1$A  and $^1$E. From there electrons decay into the $^3$A$_2$ triplet ground state with about equal rates. Due to the spin selective intersystem crossing in the excited state a stable majority population of the $\ket{0}$ can be achieved, under continuous laser pumping. 

\section*{Supplementary Note 3: Microwave Spectroscopy of the NV$^-$ Spin Ensemble and Number of Spins Estimation}

In addition to the microwave emission spectrum of the maser, we investigated the NV$^-$ spin transitions by low power microwave spectroscopy. In particular, we studied the low-field electron spin resonance (ESR) transition $\left( \ket{0}\,\mapsto\,\ket{+1} \right)$ as a function of microwave frequency and magnetic field $B_0$ and recorded the microwave reflection $S_{11}$ with a vector network analyzer (VNA). This transition is dominantly absorptive as the optically spin polarized $\ket{0}$ is energetically lying below the $\ket{+1}$ state, corresponding to a positive population difference.

\begin{figure}[b]
 \includegraphics[]{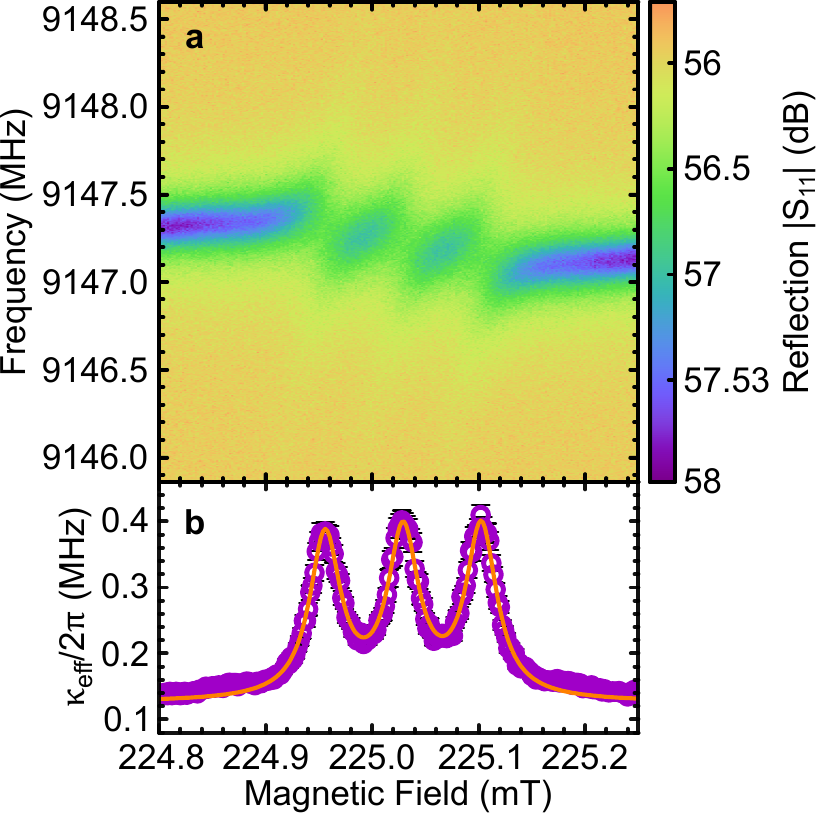}
 \caption{\textbf{Microwave reflection of low-field NV$^-$ transition.} \textbf{a} Color encoded microwave reflection $\left| S_{11} \right|$ as a function of VNA frequency and static magnetic field, at a pump rate of $695\,\second^{-1}$. \textbf{b} Effective loss rate $\kappa_{\mathrm{eff}}$ of the resonator - spin ensemble hybrid system as a function of the static magnetic field. The orange solid line is a fit of Eq.\,(\ref{eq:keff}) to the data.
 \label{S02}}
\end{figure}

Supplementary Figure \ref{S02}\,(a) shows the magnitude of the microwave reflection $\left| S_{11} \right|$ as a function of VNA frequency and static magnetic field, with an optical pump rate of $\wL = 695\,\second^{-1}$ applied to the spin ensemble. The horizontal purple line represents the microwave resonator. At the center of the field region the resonator is interrupted three times, where the resonator interacts with the NV$^-1$ spin ensemble. The splitting in three lines is due to the hyperfine interaction with the spin 1 N$^{14}$ nuclei. On resonance the resonator experiences a dispersive shift, indicating the onset of the strong coupling regime. We extract the collective coupling strength $\geff$ between the resonator and the spin ensemble to estimate the number of spins contributing to the coupling. To this end, we fit the resonator microwave reflection for each magnetic field point, using a Lorentzian model function derived from an equivalent circuit model \cite{Wang2018}.
\begin{equation}
    \left| S_{11} \right| = A \left| \frac{k - 1 - iQ_\mathrm{int}\left( \frac{\omega}{\omega_\mathrm{res}} - \frac{\omega_\mathrm{res}}{\omega} \right)}{k + 1 + iQ_\mathrm{int}\left( \frac{\omega}{\omega_\mathrm{res}} - \frac{\omega_\mathrm{res}}{\omega} \right)} \right|.
    \label{eq:lorentz}
\end{equation}
Far detuned from the spin ensemble, we can extract the internal quality factor $Q_{\mathrm{int}}$ and via the relation for the resonator coupling $k = Q_\mathrm{int} / Q_\mathrm{ext}$ the external Q-factor $Q_\mathrm{ext}$ and hence the loaded quality factor $\QL$, as well as the resonator frequency $\wc$ and its loss rate $\kappa_0$ \cite{Wang2018}. Over the full magnetic field range the loss rate changes, represented by an effective loss rate $\kappa_{\mathrm{eff}}$ and shown in Supplementary Figure \ref{S02}\,(b). $\kappa_{\mathrm{eff}}$ shows three prominent peaks at the NV$^-$ resonances, where the increased loss rate results from the hybridization of the two systems, leading to an average between resonator and spin ensemble loss rates $\kappa_0$ and $\gamma$, respectively. Using a coupled harmonic resonator model, $\kappa_{\mathrm{eff}}$ can be described using \cite{Khan2021}
\begin{equation}
    \keff = \kappa_0 + \sum_{n=-1}^{+1} \frac{g_{\mathrm{eff},n}^2 \gamma}{\Delta_n^2 + \gamma^2}.
    \label{eq:keff}
\end{equation}
The sum is over all three hyperfine split ESR transitions with $n = -1, 0, +1$ and $\Delta_n = \frac{g_{\mathrm{NV}}\mu_{\mathrm{B}}}{\hbar} \left( B_0 - B_{\mathrm{ESR}, n} \right)$ gives the detuning from the individual ESR resonances. We use the same spin loss rate for all transitions and find $\gamma /2\pi = 530.8\,\kilo\hertz \pm 7.7\,\kilo\hertz$. The collective coupling strength is the strongest for the lowest field transition and weakest for the highest field transition, which we attribute to some degree of saturation of the spin ensemble while sweeping the magnetic field over the transitions. The highest collective coupling achieved is $357.8\,\kilo\hertz \pm 2.8\,\kilo\hertz$, pumping the NV$^-$ with $695\,\second^{-1}$. 

To amount for the microwave saturation we calculate the excitation rate due to the microwave probe tone. The input microwave signal is heavily attenuated to $-55.7\,\deci\bel\milli$ at the input of the microwave cavity to limit saturation of the spin transition. The microwave signal is further attenuated by the insertion loss to the fully undercoupled resonator, given by $IL = -10\log\frac{4k}{(1 + k)^2}$ \cite{Jaynes1945, Moreno1948}. Here, k is the resonator coupling coefficient and is determined by fitting the resonator reflection signal to $0.155 \pm 0.002$ \cite{Wang2018}. This results in an insertion loss of $3.33\,\deci\bel\milli$, leading to a final microwave power in the resonator of about $-59\,\deci\bel\milli$. This corresponds to an average number of photons $\left< n \right>$ of $2.87 \times 10^8$ inside the resonator with a loaded quality factor of $\QL = 33,500$. The rate of excitation as well as for relaxation if given in SI-units by \cite{Castner1959}
\begin{equation}
\begin{split}
    \left( \frac{dn}{dt} \right)_{\mathrm{MW}} &= \frac{1}{4} n \left( \frac{g_{\mathrm{NV}}\mu_{\mathrm{B}}}{\hbar} \right)^2 B_1^2 g \left(\omega - \omega_{\mathrm{ESR}} \right), \\
    \left( \frac{dn}{dt} \right)_{\mathrm{T1}} &= \frac{n - n_0}{T_1}.
\end{split}
\end{equation}
Here, $g_{\mathrm{NV}}$ gives the NV$^-$ g-factor, $\mu_{\mathrm{B}}$ the Bohr magneton, $h$ the Planck constant, $T_1$ being the longitudinal relaxation time, $n$ and $n_0$ are the relative spin population differences under optical pumping and pure Boltzmann distribution without pumping, respectively, both normalized by the total number of spins. The additional factor of $\sfrac{1}{4}$ at the microwave excitation rate takes into account that the microwave energy can be absorbed by only one of the four NV$^-$ defect sub-sets. $g \left(\omega - \omega_{\mathrm{ESR}} \right)$ gives the linewidth function of the homogeneously broadened spin packets in the sample for which we assume a Lorentzian lineshape \cite{Castner1959}
\begin{equation}
    g \left(\omega - \wc \right) = \frac{T_2}{\pi} \frac{1}{1 + T_2^2 \left(\omega - \omega_{\mathrm{ESR}} \right)^2},
\end{equation}
where the spin-spin relaxation time is given by $T_2$ and $\left(\omega - \wc \right)$ gives the detuning from the resonance frequency of the spin transition $\omega_{\mathrm{ESR}}$. 
We estimate the rate of excitation via an averaged $B_{\mathrm{1,avg}}$ field strength, determined by finite element simulations of the sapphire ring resonator \cite{Oskooi2010} to $B_{\mathrm{1,avg}} = 1.5 \times 10^{-7}\,\tesla$. On resonance and with $T_2 = 25\,\micro\second$, determined by conventional pulsed ESR, we estimate an excitation rate of $93.0\,\second^{-1}$. At maximal laser pump rates, $T_1 = 1.5\,\milli\second$ (see previous section) and the rate of relaxation results in $45.9\,\second^{-1}$. The population difference $n$ is calculated for an optical pump rate of $695\,\second^{-1}$, following the rate equation formalism presented in \onlinecite{Sherman2021}.
With the rate of excitation being about twice as high as the relaxation rate the spins are partially saturated. Comparing the calculated spin populations with and without a microwave drive we find that the microwave drive is reducing the population difference of the $\ket{0}\,\mapsto\,\ket{+1}$ transition by $19.1\,\%$. We scale the extracted value for the collective coupling $\geff$ to remove the effects of saturation and yield $\geff/2\pi = 390\,\kilo\hertz \pm 3.1\,\kilo\hertz$.

From the finite element simulations of the resonator we are able to determine an average single spin-photon coupling rate $g_0/2\pi = 0.244\,\hertz$. Assuming a homogeneous $B_{\mathrm{1}}$-field over the diamond sample, and consequently a homogeneous distribution of $g_0$, the collective coupling is given by $\geff = g_0\sqrt{S_\mathrm{z}}$, with $S_\mathrm{z}$ being the population difference between the two spin states. With the un-saturated value for $\geff$ we find $S_\mathrm{z} = 2.57\,\times\,10^{12}$. To find the total number of spins for a given hyperfine transition and single NV$^-$ defect axis, we scale $S_\mathrm{z}$ up to $100\,\%$ spin polarization. At a pump rate of $695\,\second^{-1}$ we calculate a normalized population difference of $0.07$ between $\left( \ket{0}\,\mapsto\,\ket{+1} \right)$. Together, we find a total number of spins per hyperfine transition and per NV$^-$ defect axis to $N = 3.67\,\times\,10^{13}$. To get a good agreement of the threshold equation with the experimental data, we require to adjust the number of spins by a factor of $0.63$, leading to a final number of spins of $2.32\,\times\,10^{13}$. An accurate determination of the absolute number of spins in a given sample is typically challenging and often only accurate to the order of magnitude. Having estimated $N$ within a third to arrive at an excellent agreement with the experimental data corroborates our method to determine $N$.

\section*{Supplementary Note 4: Microwave Cavity and Resonator Design Details}

\begin{figure}[t]
 \includegraphics[scale=0.95]{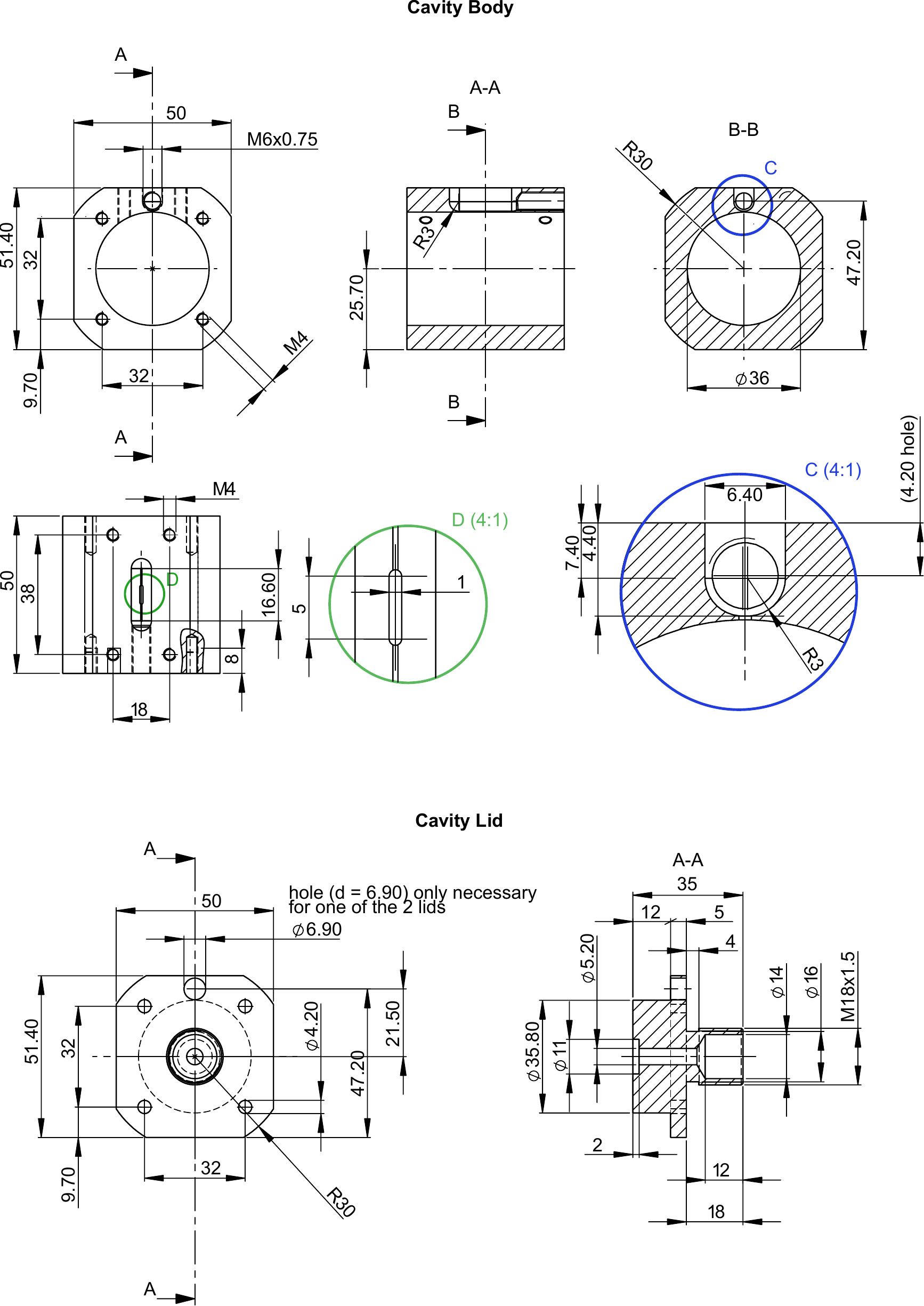}
 \caption{\textbf{Microwave Cavity Technical Drawing.} The technical drawing shows the dimensions of the gold platted copper cavity.
 \label{S03}}
\end{figure}

As described in the main text, the cavity's purpose is to suppress the radiation losses of the dielectric sapphire ring resonator. The dimensions are chosen such that the quality factor of the resonator is as large as possible, while still being able to be fitted inside of the poles of a normal conducting magnet system. Supplementary Figure \ref{S03} shows the technical drawing of the cavity, including all relevant dimensions. The body of the cavity is made of OFHC copper and is gold platted to protect the copper surfaces from oxidation. Prior gold platting a silver adhesion layer 1-2\,$\mu$m thickness was put on the copper followed by a 0.2-0.3\,$\mu$m gold layer. Since the conductivity of gold is lower than of pristine OFHC copper the thickness of the gold layer was kept minimal. The cavity is coupled via an iris to a standard X-band waveguide. 
The sapphire dielectric ring resonator is of cylindrical shape with an outer diameter of $10\,\milli\meter$, an inner diameter of $5.1\,\milli\meter$ and a height of $6\,\milli\meter$. We determine the mode volume of the resonator to $V_\mathrm{m} = 5.1 \times 10^{-8} \meter^3$, via finite element simulations \cite{Oskooi2010}. Together with the volume of the diamond of about $V_\mathrm{dia} = 10^{-8} \meter^3$ the filling factor of the resonator with diamond sample is given by $\eta = \sfrac{V_\mathrm{dia}}{V_\mathrm{m}} = 0.2$.


\begin{thebibliography}{31}
\makeatletter
\providecommand \@ifxundefined [1]{%
 \@ifx{#1\undefined}
}%
\providecommand \@ifnum [1]{%
 \ifnum #1\expandafter \@firstoftwo
 \else \expandafter \@secondoftwo
 \fi
}%
\providecommand \@ifx [1]{%
 \ifx #1\expandafter \@firstoftwo
 \else \expandafter \@secondoftwo
 \fi
}%
\providecommand \natexlab [1]{#1}%
\providecommand \enquote  [1]{``#1''}%
\providecommand \bibnamefont  [1]{#1}%
\providecommand \bibfnamefont [1]{#1}%
\providecommand \citenamefont [1]{#1}%
\providecommand \href@noop [0]{\@secondoftwo}%
\providecommand \href [0]{\begingroup \@sanitize@url \@href}%
\providecommand \@href[1]{\@@startlink{#1}\@@href}%
\providecommand \@@href[1]{\endgroup#1\@@endlink}%
\providecommand \@sanitize@url [0]{\catcode `\\12\catcode `\$12\catcode
  `\&12\catcode `\#12\catcode `\^12\catcode `\_12\catcode `\%12\relax}%
\providecommand \@@startlink[1]{}%
\providecommand \@@endlink[0]{}%
\providecommand \url  [0]{\begingroup\@sanitize@url \@url }%
\providecommand \@url [1]{\endgroup\@href {#1}{\urlprefix }}%
\providecommand \urlprefix  [0]{URL }%
\providecommand \Eprint [0]{\href }%
\providecommand \doibase [0]{http://dx.doi.org/}%
\providecommand \selectlanguage [0]{\@gobble}%
\providecommand \bibinfo  [0]{\@secondoftwo}%
\providecommand \bibfield  [0]{\@secondoftwo}%
\providecommand \translation [1]{[#1]}%
\providecommand \BibitemOpen [0]{}%
\providecommand \bibitemStop [0]{}%
\providecommand \bibitemNoStop [0]{.\EOS\space}%
\providecommand \EOS [0]{\spacefactor3000\relax}%
\providecommand \BibitemShut  [1]{\csname bibitem#1\endcsname}%
\let\auto@bib@innerbib\@empty

\bibitem{Gordon1955}
\bibinfo{author}{Gordon, J.~P.}, \bibinfo{author}{Zeiger, H.~J.} \&
  \bibinfo{author}{Townes, C.~H.}
\newblock \bibinfo{title}{The maser---new type of microwave amplifier,
  frequency standard, and spectrometer}.
\newblock \emph{\bibinfo{journal}{Physical Review}}
  \textbf{\bibinfo{volume}{99}}, \bibinfo{pages}{1264--1274}
  (\bibinfo{year}{1955}).

\bibitem{Reid1973}
\bibinfo{author}{Reid, M.}, \bibinfo{author}{Clauss, R.},
  \bibinfo{author}{Bathker, D.} \& \bibinfo{author}{Stelzried, C.}
\newblock \bibinfo{title}{Low-noise microwave receiving systems in a worldwide
  network of large antennas}.
\newblock \emph{\bibinfo{journal}{Proceedings of the IEEE}}
  \textbf{\bibinfo{volume}{61}}, \bibinfo{pages}{1330--1335}
  (\bibinfo{year}{1973}).

\bibitem{Makhov1958}
\bibinfo{author}{Makhov, G.}, \bibinfo{author}{Kikuchi, C.},
  \bibinfo{author}{Lambe, J.} \& \bibinfo{author}{Terhune, R.~W.}
\newblock \bibinfo{title}{Maser action in ruby}.
\newblock \emph{\bibinfo{journal}{Physical Review}}
  \textbf{\bibinfo{volume}{109}}, \bibinfo{pages}{1399--1400}
  (\bibinfo{year}{1958}).

\bibitem{Goldenberg1960}
\bibinfo{author}{Goldenberg, H.~M.}, \bibinfo{author}{Kleppner, D.} \&
  \bibinfo{author}{Ramsey, N.~F.}
\newblock \bibinfo{title}{Atomic hydrogen maser}.
\newblock \emph{\bibinfo{journal}{Physical Review Letters}}
  \textbf{\bibinfo{volume}{5}}, \bibinfo{pages}{361--362}
  (\bibinfo{year}{1960}).

\bibitem{Moi1983}
\bibinfo{author}{Moi, L.} \emph{et~al.}
\newblock \bibinfo{title}{Rydberg-atom masers. i. a theoretical and
  experimental study of super-radiant systems in the millimeter-wave domain}.
\newblock \emph{\bibinfo{journal}{Physical Review A}}
  \textbf{\bibinfo{volume}{27}}, \bibinfo{pages}{2043--2064}
  (\bibinfo{year}{1983}).

\bibitem{Oxborrow_Nature2012}
\bibinfo{author}{Oxborrow, M.}, \bibinfo{author}{Breeze, J.~D.} \&
  \bibinfo{author}{Alford, N.~M.}
\newblock \bibinfo{title}{Room-temperature solid-state maser}.
\newblock \emph{\bibinfo{journal}{Nature}} \textbf{\bibinfo{volume}{488}},
  \bibinfo{pages}{353--356} (\bibinfo{year}{2012}).

\bibitem{Breeze_NComms2015}
\bibinfo{author}{Breeze, J.} \emph{et~al.}
\newblock \bibinfo{title}{Enhanced magnetic purcell effect in room-temperature
  masers}.
\newblock \emph{\bibinfo{journal}{Nature Communications}}
  \textbf{\bibinfo{volume}{6}}, \bibinfo{pages}{6215} (\bibinfo{year}{2015}).

\bibitem{Salvadori2017}
\bibinfo{author}{Salvadori, E.} \emph{et~al.}
\newblock \bibinfo{title}{Nanosecond time-resolved characterization of a
  pentacene-based room-temperature maser}.
\newblock \emph{\bibinfo{journal}{Scientific Reports}}
  \textbf{\bibinfo{volume}{7}}, \bibinfo{pages}{41836} (\bibinfo{year}{2017}).

\bibitem{Jin_NComms2015}
\bibinfo{author}{Jin, L.} \emph{et~al.}
\newblock \bibinfo{title}{Proposal for a room-temperature diamond maser}.
\newblock \emph{\bibinfo{journal}{Nature Communications}}
  \textbf{\bibinfo{volume}{6}}, \bibinfo{pages}{8251} (\bibinfo{year}{2015}).

\bibitem{Breeze_Nature2018}
\bibinfo{author}{Breeze, J.~D.}, \bibinfo{author}{Salvadori, E.},
  \bibinfo{author}{Sathian, J.}, \bibinfo{author}{Alford, N.~M.} \&
  \bibinfo{author}{Kay, C. W.~M.}
\newblock \bibinfo{title}{Continuous-wave room-temperature diamond maser}.
\newblock \emph{\bibinfo{journal}{Nature}} \textbf{\bibinfo{volume}{555}},
  \bibinfo{pages}{493--496} (\bibinfo{year}{2018}).

\bibitem{Sherman2022}
\bibinfo{author}{Sherman, A.} \emph{et~al.}
\newblock \bibinfo{title}{Diamond-based microwave quantum amplifier}.
\newblock \emph{\bibinfo{journal}{Science Advances}}
  \textbf{\bibinfo{volume}{8}}, \bibinfo{pages}{eade6527}
  (\bibinfo{year}{2022}).

\bibitem{Koppenhofer2022}
\bibinfo{author}{Koppenh\"ofer, M.}, \bibinfo{author}{Groszkowski, P.},
  \bibinfo{author}{Lau, H.-K.} \& \bibinfo{author}{Clerk, A.}
\newblock \bibinfo{title}{Dissipative superradiant spin amplifier for enhanced
  quantum sensing}.
\newblock \emph{\bibinfo{journal}{PRX Quantum}} \textbf{\bibinfo{volume}{3}},
  \bibinfo{pages}{030330} (\bibinfo{year}{2022}).

\bibitem{Jiang2022_amp}
\bibinfo{author}{Jiang, M.} \emph{et~al.}
\newblock \bibinfo{title}{Floquet spin amplification}.
\newblock \emph{\bibinfo{journal}{Physical Review Letters}}
  \textbf{\bibinfo{volume}{128}}, \bibinfo{pages}{233201}
  (\bibinfo{year}{2022}).

\bibitem{Wu2022}
\bibinfo{author}{Wu, H.} \emph{et~al.}
\newblock \bibinfo{title}{Enhanced quantum sensing with room-temperature
  solid-state masers}.
\newblock \emph{\bibinfo{journal}{Science Advances}}
  \textbf{\bibinfo{volume}{8}}, \bibinfo{pages}{eade1613}
  (\bibinfo{year}{2022}).

\bibitem{Gottscholl2022}
\bibinfo{author}{Gottscholl, A.} \emph{et~al.}
\newblock \bibinfo{title}{Superradiance of spin defects in silicon carbide for
  maser applications}.
\newblock \emph{\bibinfo{journal}{Frontiers in Photonics}}
  \textbf{\bibinfo{volume}{3}} (\bibinfo{year}{2022}).

\bibitem{Jiang2022}
\bibinfo{author}{Jiang, M.}, \bibinfo{author}{Su, H.}, \bibinfo{author}{Wu,
  Z.}, \bibinfo{author}{Peng, X.} \& \bibinfo{author}{Budker, D.}
\newblock \bibinfo{title}{Floquet maser}.
\newblock \emph{\bibinfo{journal}{Science Advances}}
  \textbf{\bibinfo{volume}{7}}, \bibinfo{pages}{eabe0719}
  (\bibinfo{year}{2022}).

\bibitem{Shtin2009}
\bibinfo{author}{Shtin, N.~A.}, \bibinfo{author}{Romero, J. M.~L.} \&
  \bibinfo{author}{Prokhorov, E.}
\newblock \bibinfo{title}{Theory of fundamental microwave absorption in
  sapphire ($\alpha$-al2o3)}.
\newblock \emph{\bibinfo{journal}{Journal of Applied Physics}}
  \textbf{\bibinfo{volume}{106}}, \bibinfo{pages}{104115}
  (\bibinfo{year}{2009}).

\bibitem{Accatino1994}
\bibinfo{author}{Accatino, L.} \& \bibinfo{author}{Bertin, G.}
\newblock \bibinfo{title}{Design of coupling irises between circular cavities
  by modal analysis}.
\newblock \emph{\bibinfo{journal}{IEEE Transactions on Microwave Theory and
  Techniques}} \textbf{\bibinfo{volume}{42}}, \bibinfo{pages}{1307--1313}
  (\bibinfo{year}{1994}).

\bibitem{supplement}
\bibinfo{note}{For additional information see: Supplementary Material: Maser
  Threshold Characterization by Resonator Q-Factor Tuning}.

\bibitem{Robledo_NJP2011}
\bibinfo{author}{Robledo, L.}, \bibinfo{author}{Bernien, H.},
  \bibinfo{author}{Sar, T. v.~d.} \& \bibinfo{author}{Hanson, R.}
\newblock \bibinfo{title}{Spin dynamics in the optical cycle of single
  nitrogen-vacancy centres in diamond}.
\newblock \emph{\bibinfo{journal}{New Journal of Physics}}
  \textbf{\bibinfo{volume}{13}}, \bibinfo{pages}{025013}
  (\bibinfo{year}{2011}).

\bibitem{Kai-Mei2009}
\bibinfo{author}{Fu, K.-M.~C.} \emph{et~al.}
\newblock \bibinfo{title}{Observation of the dynamic jahn-teller effect in the
  excited states of nitrogen-vacancy centers in diamond}.
\newblock \emph{\bibinfo{journal}{Physical Review Letters}}
  \textbf{\bibinfo{volume}{103}}, \bibinfo{pages}{256404}
  (\bibinfo{year}{2009}).

\bibitem{Doherty2013}
\bibinfo{author}{Doherty, M.~W.} \emph{et~al.}
\newblock \bibinfo{title}{The nitrogen-vacancy colour centre in diamond}.
\newblock \emph{\bibinfo{journal}{Physics Reports}}
  \textbf{\bibinfo{volume}{528}}, \bibinfo{pages}{1--45}
  (\bibinfo{year}{2013}).

\bibitem{Mrozek2015}
\bibinfo{author}{Mrózek, M.} \emph{et~al.}
\newblock \bibinfo{title}{Longitudinal spin relaxation in nitrogen-vacancy
  ensembles in diamond}.
\newblock \emph{\bibinfo{journal}{EPJ Quantum Technology}}
  \textbf{\bibinfo{volume}{2}}, \bibinfo{pages}{22} (\bibinfo{year}{2015}).

\bibitem{Jeong2017}
\bibinfo{author}{Jeong, K.} \emph{et~al.}
\newblock \bibinfo{title}{Understanding the magnetic resonance spectrum of
  nitrogen vacancy centers in an ensemble of randomly oriented nanodiamonds}.
\newblock \emph{\bibinfo{journal}{Journal of Physical Chemistry C}}
  \textbf{\bibinfo{volume}{121}}, \bibinfo{pages}{21057--21061}
  (\bibinfo{year}{2017}).

\bibitem{Goeppl2008}
\bibinfo{author}{G\"{o}ppl, M.} \emph{et~al.}
\newblock \bibinfo{title}{Coplanar waveguide resonators for circuit quantum
  electrodynamics}.
\newblock \emph{\bibinfo{journal}{Journal of Applied Physics}}
  \textbf{\bibinfo{volume}{104}}, \bibinfo{pages}{--} (\bibinfo{year}{2008}).

\bibitem{Kolobov1993}
\bibinfo{author}{Kolobov, M.~I.}, \bibinfo{author}{Davidovich, L.},
  \bibinfo{author}{Giacobino, E.} \& \bibinfo{author}{Fabre, C.}
\newblock \bibinfo{title}{Role of pumping statistics and dynamics of atomic
  polarization in quantum fluctuations of laser sources}.
\newblock \emph{\bibinfo{journal}{Physical Review A}}
  \textbf{\bibinfo{volume}{47}}, \bibinfo{pages}{1431--1446}
  (\bibinfo{year}{1993}).

\bibitem{Haroche2013}
\bibinfo{author}{Haroche, S.} \& \bibinfo{author}{Raimond, J.-M.}
\newblock \emph{\bibinfo{title}{Exploring the quantum. Atoms, cavities and
  \newline photons. Reprint of the 2006 hardback ed.}}
  (\bibinfo{publisher}{Oxford Graduate Texts. Oxford: Oxford University
  Press.}, \bibinfo{year}{2013}).

\bibitem{Oskooi2010}
\bibinfo{author}{Oskooi, A.~F.} \emph{et~al.}
\newblock \bibinfo{title}{Meep: A flexible free-software package for
  electromagnetic simulations by the fdtd method}.
\newblock \emph{\bibinfo{journal}{Computer Physics Communications}}
  \textbf{\bibinfo{volume}{181}}, \bibinfo{pages}{687--702}
  (\bibinfo{year}{2010}).

\bibitem{Sherman2021}
\bibinfo{author}{Sherman, A.}, \bibinfo{author}{Buchbinder, L.},
  \bibinfo{author}{Ding, S.} \& \bibinfo{author}{Blank, A.}
\newblock \bibinfo{title}{Performance analysis of diamond-based masers}.
\newblock \emph{\bibinfo{journal}{Journal of Applied Physics}}
  \textbf{\bibinfo{volume}{129}}, \bibinfo{pages}{144503}
  (\bibinfo{year}{2021}).

\bibitem{Jarmola2012}
\bibinfo{author}{Jarmola, A.}, \bibinfo{author}{Acosta, V.~M.},
  \bibinfo{author}{Jensen, K.}, \bibinfo{author}{Chemerisov, S.} \&
  \bibinfo{author}{Budker, D.}
\newblock \bibinfo{title}{Temperature- and magnetic-field-dependent
  longitudinal spin relaxation in nitrogen-vacancy ensembles in diamond}.
\newblock \emph{\bibinfo{journal}{Physical Review Letters}}
  \textbf{\bibinfo{volume}{108}}, \bibinfo{pages}{197601}
  (\bibinfo{year}{2012}).

\bibitem{Wang2018}
\bibinfo{author}{Wang, P.} \emph{et~al.}
\newblock \bibinfo{title}{Novel method to measure unloaded quality factor of
  resonant cavities at room temperature}.
\newblock \emph{\bibinfo{journal}{Nuclear Science and Techniques}}
  \textbf{\bibinfo{volume}{29}}, \bibinfo{pages}{50} (\bibinfo{year}{2018}).

\end{thebibliography}
\end{document}